\documentclass{ecai} 

\usepackage{latexsym}
\usepackage{amssymb}
\usepackage{amsmath}
\usepackage{amsthm}
\usepackage{booktabs}
\usepackage{enumitem}
\usepackage{graphicx}
\usepackage{color}
\usepackage{caption}
\usepackage{subcaption}

\makeatletter
\def\blfootnote{\gdef\@thefnmark{}\@footnotetext}
\makeatother

\newcommand{\BibTeX}{B\kern-.05em{\sc i\kern-.025em b}\kern-.08em\TeX}

\begin{document}


\begin{frontmatter}


\paperid{123} 

\title{Understanding Language Modeling Paradigm Adaptations in Recommender Systems: Lessons Learned and Open Challenges}


\author[A]{\fnms{Lemei}~\snm{Zhang}}
\author[A]{\fnms{Peng}~\snm{liu}}
\author[B]{\fnms{Yashar}~\snm{Deldjoo}} 
\author[C]{\fnms{Yong}~\snm{Zheng}}
\author[A]{\fnms{Jon Atle}~\snm{Gulla}}

\address[A]{Norwegian University of Science and Technology, Norway}
\address[B]{Polytechnic University of Bari, Italy}
\address[C]{Illinois Institute of Technology, USA}


\begin{abstract}
The emergence of Large Language Models (LLMs) has achieved tremendous success in the field of Natural Language Processing owing to diverse training paradigms that empower LLMs to effectively capture intricate linguistic patterns and semantic representations. In particular, the recent "pre-train, prompt and predict" training paradigm has attracted significant attention as an approach for learning generalizable models with limited labeled data. In line with this advancement, these training paradigms have recently been adapted to the recommendation domain and are seen as a promising direction in both academia and industry. This half-day tutorial aims to provide a thorough understanding of extracting and transferring knowledge from pre-trained models learned through different training paradigms to improve recommender systems from various perspectives,  such as generality, sparsity, effectiveness and trustworthiness. In this tutorial, we first introduce the basic concepts and a generic architecture of the language modeling paradigm for recommendation purposes. Then, we focus on recent advancements in adapting LLM-related training strategies and  optimization objectives for different recommendation tasks. After that, we will systematically introduce ethical issues in LLM-based recommender systems and discuss possible approaches to assessing and mitigating them. We will also summarize the relevant datasets, evaluation metrics, and an empirical study on the recommendation performance of training paradigms. Finally, we will conclude the tutorial with a discussion of open challenges and future directions.
\end{abstract}

\end{frontmatter}


\section{Introduction}
Recommender Systems\blfootnote{\emph{ECAI'24, 19-24 October, 2024, Santiago de Compostela, Spain}} (RSs) focus users on a small selection of items from a much larger catalog, alleviating information overload and boosting sales for Internet retailers. RSs have become an essential part of the large Internet today, driving up to 50-80\% of sales or consumed content, due to their efficacy \cite{deldjoo2023review}. With the thriving of Large Language Models (LLMs) and their remarkable success in Natural Language Processing (NLP), the adaptation of their training paradigms for recommender systems has attracted significant attention in both academia and industry. Specially designed LLMs for recommender systems enhance traditional RSs by extracting nuanced textual representations and leveraging extensive external knowledge to understand user preferences in a more nuanced way, thereby aligning with the goal of delivering personalized, context-aware recommendations.

The advantages of LLMs in recommender systems can also be their disadvantage; their training on vast, unregulated Internet data may encode biases against specific races, genders, or brands, leading to potentially unfair recommendations. For instance, an LLM trained mainly on data from leading e-commerce platforms may favor well-known brands over niche or upcoming ones, and inherent biases related to gender, race, or even specific genres and years could subtly but significantly influence recommendations \cite{deldjoo2024understanding}.

\begin{figure*}[t]
     \centering
     \begin{subfigure}[b]{0.46\textwidth}
         \centering
         \includegraphics[width=\textwidth]{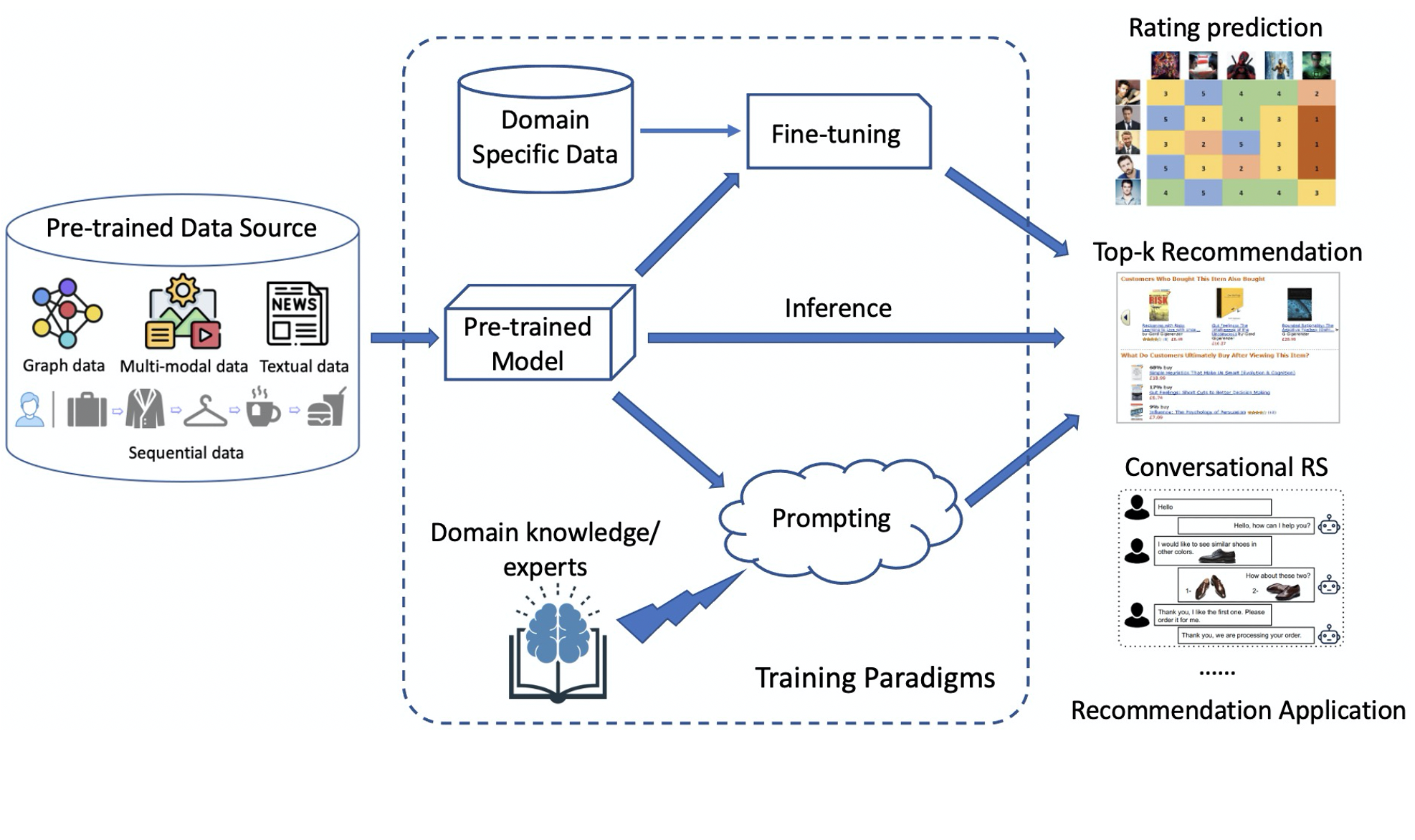}
         \caption{}
         \label{fig:architecture_rs}
     \end{subfigure}
     \hfill
     \begin{subfigure}[b]{0.53\textwidth}
         \centering
         \includegraphics[width=\textwidth]{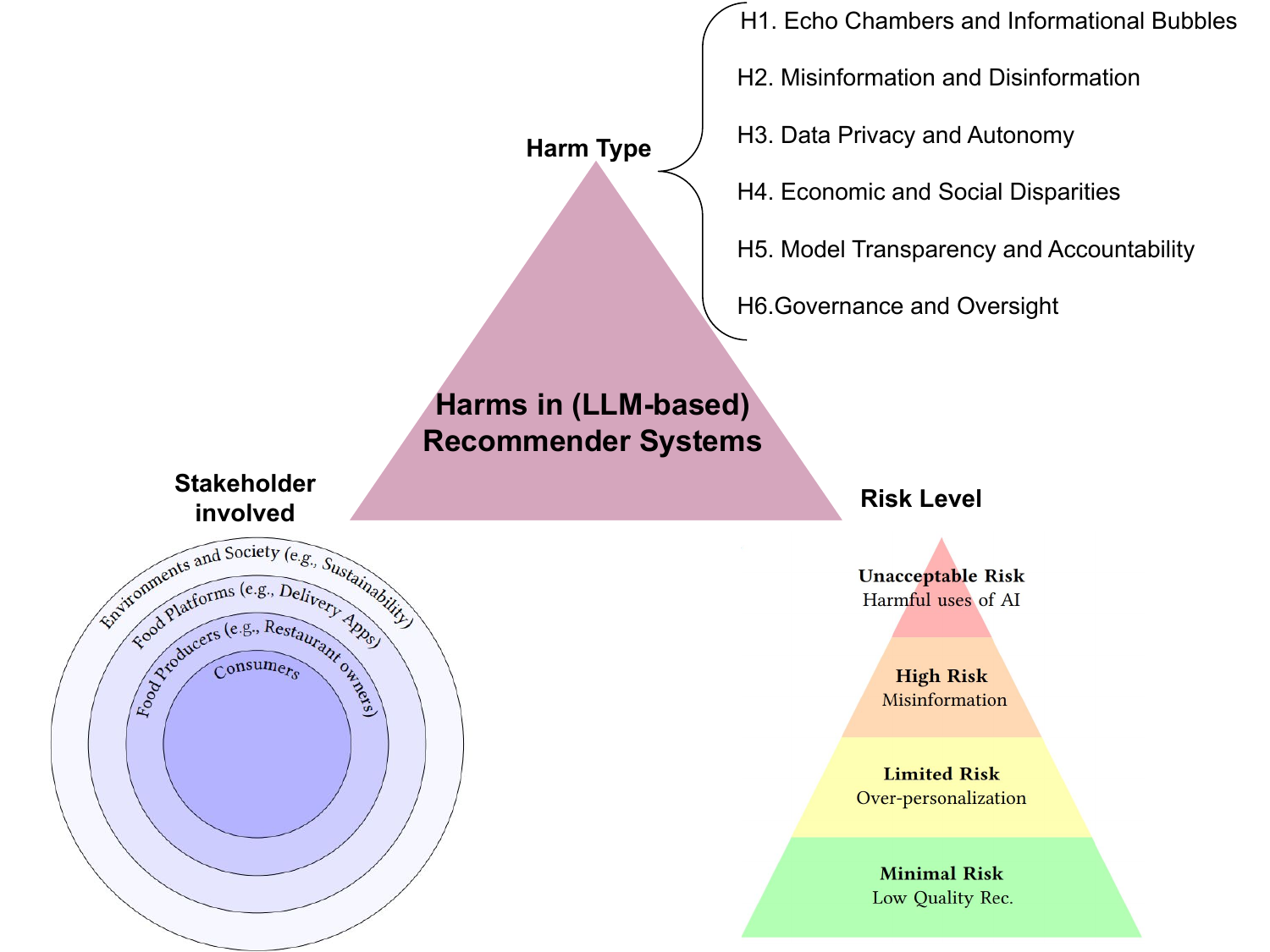}
         \caption{}
         \label{fig:harms_rs}
     \end{subfigure}
     \vspace{8mm}
        \caption{(a) A generic architecture of language modeling paradigm for RSs. (b) Exploring Harms and Ethical Considerations in LLM-based RSs. Please note that several topics discussed here also intersect with those in traditional RSs.}
        \label{fig:llm_rs}
\vspace{8mm}
\end{figure*}

This tutorial aims to provide a comprehensive and in-depth exploration of the complexities involved in adapting language modeling paradigms to improve the performance and ethical considerations of recommender systems. It aims to present various training strategies and optimization objectives that facilitate the extraction and transfer of knowledge from pre-trained models to recommendation tasks, offering insights into theoretical foundations, practical implications, open challenges, and emerging trends in this rapidly evolving and impactful domain. Moreover, the tutorial can serve as a platform for exchanging ideas, sharing best practices, and encouraging collaborations among experts in the field, ultimately driving advancements and innovation in the adaptation of language modeling paradigms for RSs. The scope of the tutorial encompasses a thorough understanding of the following five key aspects:

\begin{itemize}[leftmargin=*]
\item \textbf{Basic Concepts and Architecture:} We briefly introduce the application context on RS to the audience and enumerate various tasks and areas where RS help users/businesses etc. We will also clarify basic concepts and a generic architecture (see Figure \ref{fig:llm_rs}(a)) of LLM-based RSs to provide the audience with a foundational understanding of the topic.
\item \textbf{Adaptation of LLM-Related Training Paradigms:} Building upon the basics, we retrospect the recent advancements in adapting LLM-related training strategies \citep{geng2022path,xiao2022training,hou2022towards,liu2023graph,wang2022towards,deng2023unified,xin2022rethinking} and optimization objectives \citep{bao2020unilmv2,liu2019roberta,zhao2022resetbert4rec,wang2023curriculum,wu2022mm} for various recommendation tasks. As illustrated in Figure \ref{fig:llm_rs}(a), there are mainly two classes of training paradigms: pre-train, fine-tune paradigm, and prompt learning paradigm. Each class is further classified into subclasses w.r.t. different training efforts on different parts of the recommendation model. It explores how these paradigms can be modified and optimized to accommodate specific nuances and challenges inherent in RSs, including issues related to generality, sparsity, and effectiveness.
\item \textbf{Ethical Considerations in LLM-based RSs:} Fundamentally, trust in recommender systems is correlated with the risks or harms they might pose. In addressing ethical concerns and trustworthiness, we examine three key questions as shown in Figure \ref{fig:llm_rs}(b): (1) \emph{What types of harm might occur?} While harm is present in traditional, non-LLM-based recommender systems, it is important to recognize that LLM-based RS could potentially exacerbate them. We have identified six types of potential harms in recommender systems \citep{weidinger2021ethical}: echo chambers and information bubbles, misinformation and disinformation, data privacy and autonomy, economic and social disparities, model transparency and accountability, and governance and oversight. (2) \emph{Who could be harmed?} The range of stakeholders potentially affected by harm in recommender systems extends well beyond end-users. Essentially, a multitude of stakeholders are involved, including content creators, system designers, the wider supply chain participants, and even the environment. Acknowledging the breadth of these parties is crucial to fully comprehending the possible harms that may ensue. (3) \emph{How severe are these potential harms?} Drawing inspiration from the recent European AI Act \cite{hupont2023use,european2021laying}, we suggest a categorization of risks ranging from unacceptable and high to limited and minimal, tailored to the harm's nature and intensity. Understanding these harms leads us to the fundamental question: \emph{How do we assess and mitigate them?} We start discussing possible approaches here by involving a variety of adverse indications, holding interviews with users, incorporating insights from external knowledge bases, and evaluating the impact over both the immediate and longer terms.
\item \textbf{Evaluation Framework and Empirical Studies:} We provide insights into the relevant datasets \citep{he2016ups,li2018towards,wu2020mind} and evaluation metrics from both recommendation \citep{zhang2022keep,liu2022boosting,zangerle2022evaluating} and generation \citep{xie2023factual,geng2022improving,wang2021recindial} perspectives to measure the performance of adapted RSs across various application domains. We incorporate empirical studies and real-world examples to illustrate the practical implications and benefits of adopting different training paradigms.
\item \textbf{Future Directions and Open Challenges:} We shed light on open challenges and potential future directions to help beginners and practitioners in this field to critically analyze existing limitations and explore innovative approaches to address these challenges.
\end{itemize}

This tutorial is particularly relevant to the ECAI Conference, where Language models and RSs are two rising topics in AI technologies, and the synergy between them can transform the way users interact with the Internet, such as ChatGPT and Microsoft Bing Chat. With the rapid expansion of digital platforms and the increasing demand for personalized content delivery, it becomes paramount to integrate cutting-edge recommendation strategies and control mechanisms to mitigate potential harms. The tutorial directly addresses these challenges and aligns with the conference's goal of advancing AI technologies.

\noindent\textbf{Objectives:} Our tutorial specifically serves the following obejctives:
\begin{itemize}[leftmargin=*]
    \item Introduce novices to major topics within Artificial Intelligence.
    \item Provide instruction in established but specialized AI methodologies.
    \item Motivate and explain a topic of emerging importance for AI.
\end{itemize}

Our team of presenters brings a wealth of experience and expertise to this tutorial. With a deep understanding of both LLMs and RSs, we have conducted extensive research in this emerging field, publishing papers and contributing to the development of state-of-the-art RSs. Our collective knowledge and practical experience ensure that participants will gain valuable insights not only into the current state-of-the-art but also into the potential future directions and challenges that lie ahead in the field of adapting language modeling paradigms for RSs.

\section{TUTORIAL OVERVIEW}
\subsection{Brief Schedule}
This will be a 3-hour lecture-style tutorial divided into six parts:

\begin{itemize}[leftmargin=10mm]
\item[Part 1] Introduction (20 mins)
\begin{itemize}
    \item Overview of Language Models and RSs.
    \item Overview of Language Modeling Paradigm in RSs.
\end{itemize}
\item[Part 2] Training Strategies of LLM-based RSs (45 mins)
\begin{itemize}
    \item Pre-train, fine-tune paradigm for RSs
    \subitem{Pre-train}
    \subitem{Pre-train, fine-tune the holistic model}
    \subitem{Pre-train, fine-tune partial model}
    \subitem{Pre-train, fine-tune the extra part of the model}
    \item Prompting paradigm for RSs
    \subitem{Fixed-Pretrained Model (PTM) prompt tuning}
    \subitem{Fixed-prompt PTM tuning}
    \subitem{Tuning-free prompting}
    \subitem{Prompt+PTM tuning}
\end{itemize}
\item[Part 3] Optimization Objectives of LLM-based RSs (20 mins)
\begin{itemize}
    \item Language modeling objectives to recommendation
    \item Adaptive objectives to recommendation
\end{itemize}
\item[Part 4] Ethical Issues and Trustworthiness of LLM-based RSs (45 mins)
\begin{itemize}
    \item Different harm types, stakeholders involved, and harm severity in LLM-based RSs
    \item Possible approaches to assessing and mitigating ethical issues and harms
\end{itemize}

\item[Part 5] Evaluation and Available Resources (30 mins)
\begin{itemize}
    \item Evaluation on recommendation accuracy and language perspectives
    \item Open-sourced datasets and training platforms
\end{itemize}
\item[Part 6] Summary and Future Directions (20 mins)
\end{itemize}

\subsection{Publicly Available Material}
The tutorial is based on our \textbf{two survey} papers: one has been accepted by the Transactions of the Association for Computational Linguistics (TACL)~\cite{peng2023}, and the other has been published in the User Modeling and User-Adapted Interaction (UMUAI) journal~\cite{deldjoo2023fairness_umuai}. Apart from that, some insights of our tutorial are also from \cite{weidinger2021ethical}. The slides of this tutorial along with additional information are publicly available at \url{https://github.com/SmartmediaAI/LLM_RecSys}.

\subsection{Targeted Audience}
This is an intermediate-level tutorial primarily targeting graduate students, academic researchers, and industry practitioners who have interests in adopting language modelling technologies for personalization, user modeling, and recommendation. The tutorial will also appeal to researchers and practitioners who work in broader AI/ML communities, such as NLP and Responsible Web, since we will introduce how large language models can be personalized and adapted for recommendation tasks, and discuss the responsible use of these models. A general background in NLP and Information Retrieval is sufficient, as we will introduce the basic knowledge of LLMs and RSs involved. After the tutorial, audiences can expect to gain a comprehensive understanding of the integration of LLMs into RSs, acquire the knowledge and skills necessary to design and implement appropriate LLM adaptation strategies for various recommendation tasks, evaluate LLM-based RSs using multifaceted metrics, and address ethical challenges related to bias and fairness in recommendations.

\section{BRIEF BIO OF ORGANIZERS}
\textbf{Lemei Zhang} is a postdoctoral researcher at the Norwegian Research Center for AI Innovation (NorwAI) at NTNU, Norway. Her research topics include natural language processing, recommender systems, and user modeling. Specifically, she focuses on applying data mining and natural language processing techniques to design effective algorithms that enhance the performance of recommender systems in various domains. Her current research is focusing on the development of foundational language models and their application in recommender systems. Her research has appeared in TACL, TOIS, UMUAI, RecSys, ECML-PKDD, etc., and she actively serves as a PC member or reviewer for conferences and journals such as AAAI, ACM Computing Surveys, UMUAI, and TWEB.\\

\noindent\textbf{Peng Liu} is a postdoctoral researcher at the Norwegian Research Center for AI Innovation (NorwAI) at NTNU, Norway. His research focuses on the intersection of Natural Language Processing and Recommender Systems. His primary interests lie in the areas of language modeling, sentiment analysis, and recommendation algorithms based on data streams and multimodal contexts such as text and images. His research appears in TACL, NAACL, SIGIR, RecSys, TOIS, UMUAI, ECML-PKDD, etc., and he regularly serves as a PC member or reviewer for conferences and journals such as ACL, EMNLP, NAACL, IJCAI, AAAI, TKDE, TOIS, UMUAI, etc.\\

\noindent\textbf{Yashar Deldjoo} is a tenure-track Assistant Professor at the Polytechnic University of Bari, Italy. He earned his Ph.D. with distinction in recommender systems from the Polytechnic University of Milan, Italy's best technical university. His main research is centered on the study and the development of recommender systems that are not only accurate but also  ``trustworthy'', and adhere to human norms and values. His work emphasizes model fairness, robustness, privacy, and interpretability. His recent research also studies the application of ``Generative AI and LLMs'' in Recommender Systems and ML applications \cite{nazary2023chatgpt,deldjoo2023fairness}. He has served as the lead author on three comprehensive survey papers for ACM 
CSUR \cite{deldjoo2022survey,deldjoo2023review,deldjoo2021recommender}, as well as on a notable survey on ``fairness in recommender systems'' for the UMUAI journal \cite{deldjoo2023fairness_umuai}. He regularly publishes at SIGIR, RecSys, ECIR, ECAI, CVPR, and journals, including CSUR, UMUAI, IP\&M, TKDE, and TIST. He contributed two book chapters to the 3rd Edition of ``Recommender Systems Handbook'', the best book in the RecSys community. He presented tutorials at IR/RSs venues, including WSDM, RecSys, and ECIR. Additionally, he plays an active role in the academic community, organizing notable workshops like the ACM RecSys Challenge and MediaEval, as well as IIR '21.\\

\noindent\textbf{Yong Zheng} is an Assistant Professor at the Department of Information Technology and Management within the Illinois Institute of Technology, in USA. His primary research focus lies in user modeling and recommender systems. He has contributed to various renowned conferences such as ACM RecSys, ACM UMAP, and ACM IUI as an organizing committee member. He has experience in delivering academic tutorials, such as his tutorials on topics like multi-criteria recommender systems at ACM IUI 2023, multi-objective recommender systems at ACM SIGKDD 2021, and multi-stakeholder recommender systems at ACM RecSys 2019, etc. \\

\noindent\textbf{Jon Atle Gulla} is a professor of information systems at the Norwegian University of Science and Technology since 2002 and the director of the Norwegian Research Center for AI Innovation. He has a Ph.D. in computer science from 1993 and holds three M.Sc. degrees in computer science, linguistics, and management. His research is on natural language processing and semantics in the context of recommender systems, search engines, and conversational systems. He has close to 150 international publications and has supervised around 70 MSc students, 30 PhDs and 10 Postdocs.


\bibliography{mybibfile}

\begin{thebibliography}{32}
\providecommand{\natexlab}[1]{#1}
\providecommand{\url}[1]{\texttt{#1}}
\expandafter\ifx\csname urlstyle\endcsname\relax
  \providecommand{\doi}[1]{doi: #1}\else
  \providecommand{\doi}{doi: \begingroup \urlstyle{rm}\Url}\fi

\bibitem[Bao et~al.(2020)Bao, Dong, Wei, Wang, Yang, Liu, Wang, Gao, Piao,
  Zhou, et~al.]{bao2020unilmv2}
H.~Bao, L.~Dong, F.~Wei, W.~Wang, N.~Yang, X.~Liu, Y.~Wang, J.~Gao, S.~Piao,
  M.~Zhou, et~al.
\newblock Unilmv2: Pseudo-masked language models for unified language model
  pre-training.
\newblock In \emph{International conference on machine learning}, pages
  642--652. PMLR, 2020.

\bibitem[Deldjoo(2023)]{deldjoo2023fairness}
Y.~Deldjoo.
\newblock Fairness of chatgpt and the role of explainable-guided prompts.
\newblock \emph{arXiv preprint arXiv:2307.11761}, 2023.

\bibitem[Deldjoo(2024)]{deldjoo2024understanding}
Y.~Deldjoo.
\newblock Understanding biases in chatgpt-based recommender systems: Provider
  fairness, temporal stability, and recency.
\newblock \emph{arXiv preprint arXiv:2401.10545}, 2024.

\bibitem[Deldjoo et~al.(2021)Deldjoo, Schedl, Cremonesi, and
  Pasi]{deldjoo2021recommender}
Y.~Deldjoo, M.~Schedl, P.~Cremonesi, and G.~Pasi.
\newblock Recommender systems leveraging multimedia content.
\newblock \emph{ACM Computing Surveys}, 53\penalty0 (5):\penalty0 1--38, 2021.

\bibitem[Deldjoo et~al.(2022)Deldjoo, Di~Noia, and Merra]{deldjoo2022survey}
Y.~Deldjoo, T.~Di~Noia, and F.~A. Merra.
\newblock A survey on adversarial recommender systems: from attack/defense
  strategies to generative adversarial networks.
\newblock \emph{ACM Computing Surveys}, \penalty0 (2):\penalty0 1--38, 2022.

\bibitem[Deldjoo et~al.(2023{\natexlab{a}})Deldjoo, Jannach, Bellogin, Difonzo,
  and Zanzonelli]{deldjoo2023fairness_umuai}
Y.~Deldjoo, D.~Jannach, A.~Bellogin, A.~Difonzo, and D.~Zanzonelli.
\newblock Fairness in recommender systems: research landscape and future
  directions.
\newblock \emph{User Modeling and User-Adapted Interaction}, pages 1--50,
  2023{\natexlab{a}}.

\bibitem[Deldjoo et~al.(2023{\natexlab{b}})Deldjoo, Nazary, Ramisa, Mcauley,
  Pellegrini, Bellogin, and Di~Noia]{deldjoo2023review}
Y.~Deldjoo, F.~Nazary, A.~Ramisa, J.~Mcauley, G.~Pellegrini, A.~Bellogin, and
  T.~Di~Noia.
\newblock A review of modern fashion recommender systems.
\newblock \emph{ACM Computing Surveys}, 56\penalty0 (4):\penalty0 1--37,
  2023{\natexlab{b}}.

\bibitem[Deng et~al.(2023)Deng, Zhang, Xu, Lei, Chua, and Lam]{deng2023unified}
Y.~Deng, W.~Zhang, W.~Xu, W.~Lei, T.-S. Chua, and W.~Lam.
\newblock A unified multi-task learning framework for multi-goal conversational
  recommender systems.
\newblock \emph{ACM Transactions on Information Systems}, 41\penalty0
  (3):\penalty0 1--25, 2023.

\bibitem[{European Commission}(2021)]{european2021laying}
{European Commission}.
\newblock Laying down harmonised rules on artificial intelligence (artificial
  intelligence act) and amending certain union legislative acts.
\newblock \emph{Eur Comm}, 106:\penalty0 1--108, 2021.

\bibitem[Geng et~al.(2022{\natexlab{a}})Geng, Fu, Ge, Li, De~Melo, and
  Zhang]{geng2022improving}
S.~Geng, Z.~Fu, Y.~Ge, L.~Li, G.~De~Melo, and Y.~Zhang.
\newblock Improving personalized explanation generation through visualization.
\newblock In \emph{Proceedings of the 60th Annual Meeting of the Association
  for Computational Linguistics (Volume 1: Long Papers)}, pages 244--255,
  2022{\natexlab{a}}.

\bibitem[Geng et~al.(2022{\natexlab{b}})Geng, Fu, Tan, Ge, De~Melo, and
  Zhang]{geng2022path}
S.~Geng, Z.~Fu, J.~Tan, Y.~Ge, G.~De~Melo, and Y.~Zhang.
\newblock Path language modeling over knowledge graphsfor explainable
  recommendation.
\newblock In \emph{Proceedings of the ACM Web Conference 2022}, pages 946--955,
  2022{\natexlab{b}}.

\bibitem[He and McAuley(2016)]{he2016ups}
R.~He and J.~McAuley.
\newblock Ups and downs: Modeling the visual evolution of fashion trends with
  one-class collaborative filtering.
\newblock In \emph{proceedings of the 25th international conference on world
  wide web}, pages 507--517, 2016.

\bibitem[Hou et~al.(2022)Hou, Mu, Zhao, Li, Ding, and Wen]{hou2022towards}
Y.~Hou, S.~Mu, W.~X. Zhao, Y.~Li, B.~Ding, and J.-R. Wen.
\newblock Towards universal sequence representation learning for recommender
  systems.
\newblock In \emph{Proceedings of the 28th ACM SIGKDD Conference on Knowledge
  Discovery and Data Mining}, pages 585--593, 2022.

\bibitem[Hupont et~al.(2023)Hupont, Fern{\'a}ndez-Llorca, Baldassarri, and
  G{\'o}mez]{hupont2023use}
I.~Hupont, D.~Fern{\'a}ndez-Llorca, S.~Baldassarri, and E.~G{\'o}mez.
\newblock Use case cards: a use case reporting framework inspired by the
  european ai act.
\newblock \emph{arXiv preprint arXiv:2306.13701}, 2023.

\bibitem[Li et~al.(2018)Li, Ebrahimi~Kahou, Schulz, Michalski, Charlin, and
  Pal]{li2018towards}
R.~Li, S.~Ebrahimi~Kahou, H.~Schulz, V.~Michalski, L.~Charlin, and C.~Pal.
\newblock Towards deep conversational recommendations.
\newblock \emph{Advances in neural information processing systems}, 31, 2018.

\bibitem[Liu et~al.(2023{\natexlab{a}})Liu, Zhang, and Gulla]{peng2023}
P.~Liu, L.~Zhang, and J.~A. Gulla.
\newblock Pre-train, prompt and recommendation: A comprehensive survey of
  language modelling paradigm adaptations in recommender systems.
\newblock \emph{arXiv preprint arXiv:2302.03735}, 2023{\natexlab{a}}.

\bibitem[Liu et~al.(2022)Liu, Zhu, Dai, and Wu]{liu2022boosting}
Q.~Liu, J.~Zhu, Q.~Dai, and X.~Wu.
\newblock Boosting deep ctr prediction with a plug-and-play pre-trainer for
  news recommendation.
\newblock In \emph{Proceedings of the 29th International Conference on
  Computational Linguistics}, pages 2823--2833, 2022.

\bibitem[Liu et~al.(2023{\natexlab{b}})Liu, Meng, Macdonald, and
  Ounis]{liu2023graph}
S.~Liu, Z.~Meng, C.~Macdonald, and I.~Ounis.
\newblock Graph neural pre-training for recommendation with side information.
\newblock \emph{ACM Transactions on Information Systems}, 41\penalty0
  (3):\penalty0 1--28, 2023{\natexlab{b}}.

\bibitem[Liu et~al.(2019)Liu, Ott, Goyal, Du, Joshi, Chen, Levy, Lewis,
  Zettlemoyer, and Stoyanov]{liu2019roberta}
Y.~Liu, M.~Ott, N.~Goyal, J.~Du, M.~Joshi, D.~Chen, O.~Levy, M.~Lewis,
  L.~Zettlemoyer, and V.~Stoyanov.
\newblock Roberta: A robustly optimized bert pretraining approach.
\newblock \emph{arXiv preprint arXiv:1907.11692}, 2019.

\bibitem[Nazary et~al.(2023)Nazary, Deldjoo, and Di~Noia]{nazary2023chatgpt}
F.~Nazary, Y.~Deldjoo, and T.~Di~Noia.
\newblock Chatgpt-healthprompt. harnessing the power of xai in prompt-based
  healthcare decision support using chatgpt.
\newblock \emph{XL-ML@ECAI'23}, 2023.

\bibitem[Wang et~al.(2023)Wang, Zhou, Zhao, Wang, and Wen]{wang2023curriculum}
H.~Wang, K.~Zhou, X.~Zhao, J.~Wang, and J.-R. Wen.
\newblock Curriculum pre-training heterogeneous subgraph transformer for top-n
  recommendation.
\newblock \emph{ACM Transactions on Information Systems}, 41\penalty0
  (1):\penalty0 1--28, 2023.

\bibitem[Wang et~al.(2021)Wang, Hu, Sha, Xu, Wong, and
  Jiang]{wang2021recindial}
L.~Wang, H.~Hu, L.~Sha, C.~Xu, K.-F. Wong, and D.~Jiang.
\newblock Recindial: A unified framework for conversational recommendation with
  pretrained language models.
\newblock \emph{arXiv preprint arXiv:2110.07477}, 2021.

\bibitem[Wang et~al.(2022)Wang, Zhou, Wen, and Zhao]{wang2022towards}
X.~Wang, K.~Zhou, J.-R. Wen, and W.~X. Zhao.
\newblock Towards unified conversational recommender systems via
  knowledge-enhanced prompt learning.
\newblock In \emph{Proceedings of the 28th ACM SIGKDD Conference on Knowledge
  Discovery and Data Mining}, pages 1929--1937, 2022.

\bibitem[Weidinger et~al.(2021)Weidinger, Mellor, Rauh, Griffin, Uesato, Huang,
  Cheng, Glaese, Balle, Kasirzadeh, et~al.]{weidinger2021ethical}
L.~Weidinger, J.~Mellor, M.~Rauh, C.~Griffin, J.~Uesato, P.-S. Huang, M.~Cheng,
  M.~Glaese, B.~Balle, A.~Kasirzadeh, et~al.
\newblock Ethical and social risks of harm from language models.
\newblock \emph{arXiv preprint arXiv:2112.04359}, 2021.

\bibitem[Wu et~al.(2022)Wu, Wu, Qi, Zhang, Huang, and Xu]{wu2022mm}
C.~Wu, F.~Wu, T.~Qi, C.~Zhang, Y.~Huang, and T.~Xu.
\newblock Mm-rec: Visiolinguistic model empowered multimodal news
  recommendation.
\newblock In \emph{Proceedings of the 45th International ACM SIGIR Conference
  on Research and Development in Information Retrieval}, pages 2560--2564,
  2022.

\bibitem[Wu et~al.(2020)Wu, Qiao, Chen, Wu, Qi, Lian, Liu, Xie, Gao, Wu,
  et~al.]{wu2020mind}
F.~Wu, Y.~Qiao, J.-H. Chen, C.~Wu, T.~Qi, J.~Lian, D.~Liu, X.~Xie, J.~Gao,
  W.~Wu, et~al.
\newblock Mind: A large-scale dataset for news recommendation.
\newblock In \emph{Proceedings of the 58th Annual Meeting of the Association
  for Computational Linguistics}, pages 3597--3606, 2020.

\bibitem[Xiao et~al.(2022)Xiao, Liu, Shao, Di, Middha, Wu, and
  Xie]{xiao2022training}
S.~Xiao, Z.~Liu, Y.~Shao, T.~Di, B.~Middha, F.~Wu, and X.~Xie.
\newblock Training large-scale news recommenders with pretrained language
  models in the loop.
\newblock In \emph{Proceedings of the 28th ACM SIGKDD Conference on Knowledge
  Discovery and Data Mining}, pages 4215--4225, 2022.

\bibitem[Xie et~al.(2023)Xie, Singh, McAuley, and Majumder]{xie2023factual}
Z.~Xie, S.~Singh, J.~McAuley, and B.~P. Majumder.
\newblock Factual and informative review generation for explainable
  recommendation.
\newblock In \emph{Proceedings of the AAAI Conference on Artificial
  Intelligence}, volume~37, pages 13816--13824, 2023.

\bibitem[Xin et~al.(2022)Xin, Pimentel, Karatzoglou, Ren, Christakopoulou, and
  Ren]{xin2022rethinking}
X.~Xin, T.~Pimentel, A.~Karatzoglou, P.~Ren, K.~Christakopoulou, and Z.~Ren.
\newblock Rethinking reinforcement learning for recommendation: A prompt
  perspective.
\newblock In \emph{Proceedings of the 45th International ACM SIGIR Conference
  on Research and Development in Information Retrieval}, pages 1347--1357,
  2022.

\bibitem[Zangerle and Bauer(2022)]{zangerle2022evaluating}
E.~Zangerle and C.~Bauer.
\newblock Evaluating recommender systems: survey and framework.
\newblock \emph{ACM Computing Surveys}, 55\penalty0 (8):\penalty0 1--38, 2022.

\bibitem[Zhang et~al.(2022)Zhang, Chan, Xu, Bian, Han, Deng, and
  Zheng]{zhang2022keep}
Y.~Zhang, Z.~Chan, S.~Xu, W.~Bian, S.~Han, H.~Deng, and B.~Zheng.
\newblock Keep: An industrial pre-training framework for online recommendation
  via knowledge extraction and plugging.
\newblock In \emph{Proceedings of the 31st ACM International Conference on
  Information \& Knowledge Management}, pages 3684--3693, 2022.

\bibitem[Zhao(2022)]{zhao2022resetbert4rec}
Q.~Zhao.
\newblock Resetbert4rec: A pre-training model integrating time and user
  historical behavior for sequential recommendation.
\newblock In \emph{Proceedings of the 45th international ACM SIGIR conference
  on research and development in information retrieval}, pages 1812--1816,
  2022.

\end{thebibliography}

\end{document}